\documentclass{ws-ijmpcs}

\begin{document}

\markboth{C. Sigismondi}
{Solar diameter and low frequency seeing}

%
\catchline{}{}{}{}{}
%
\title{LOW FREQUENCY SEEING AND SOLAR DIAMETER MEASUREMENTS}

\author{COSTANTINO SIGISMONDI}

\address{Sapienza University of Rome, Physics Dept., and Galileo Ferraris Institute\\
P.le Aldo Moro 5 Roma, 00185, Italy. e-mail: sigismondi@icra.it\\
University of Nice-Sophia Antipolis - Dept. Fizeau (France);\\
IRSOL, Istituto Ricerche Solari di Locarno (Switzerland);
Key Laboratory of Solar Activity, National Astronomical Observatories, Chinese Academy of Sciences, Beijing (China).\\
}

\author{XIAOFAN WANG}

\address{Key Laboratory of Solar Activity, National Astronomical Observatories, \\
Chinese Academy of Sciences, Beijing.  e-mail: wxf@nao.cas.cn}

\maketitle

\begin{history}
\received{2 Feb 2012}
\revised{Day Month Year}
\end{history}

\begin{abstract}

The action of the atmospheric seeing is blurring, image stretching and image motion. This happens even to the image of the Sun which is more than half degree wide.
Low frequency seeing components affect the solar diameter values measured either through the drift-scan or the heliometer methods.
We present evidences of image motion and stretching down to 0.001 Hz.

\end{abstract}

    
\ccode{PACS numbers: 95.75.-z; 95.75.Wx; 96.60.Bn; 96.60.-j; 96.60.Q-; 95.10.Jk, 92.60.hk,}

\section{Measuring The Solar Diameter with Drift-Scan Method}

The method consists in the use of a telescope aiming at a fixed direction, observing the drift of the solar image through the local meridian, an hourly circle or a given almucantarat (case of solar astrolabes). 
Knowing with accuracy the angular speed of the solar transit and measuring the transit time, it can give an accurate measure of the solar diameter. It should also be noted that the measured time is not affected by atmospheric refraction. Another advantage of the drift-scan method is that it is not affected by optical defects and aberrations, because both edges are observed aiming in the same direction, moreover with parallel transits\cite{sigiIAU} one can gather N observations, at rather homogeneous seeing conditions, during the time of a single transit.
The measurements of the solar diameter by meridian transit were monitored on a daily basis since 1851 at Greenwich Observatory,\cite{Gething} and at the Campidoglio (Capitol) Observatory in Rome since 1877 to 1937 (see Fig. 1).
Despite the advantages of this method, independent measurements seem to be inconsistent.\cite{Wittmann,Wittmann2} This discrepancy could be due to the importance of seeing global motions acting for periods longer than the typical time of the transit (about 2 minutes). 
To evaluate this hypothesis, we perform a survey on the low frequency regions of the seeing power spectrum.

\begin{figure}
\centerline{\includegraphics[width=1\textwidth,clip=]{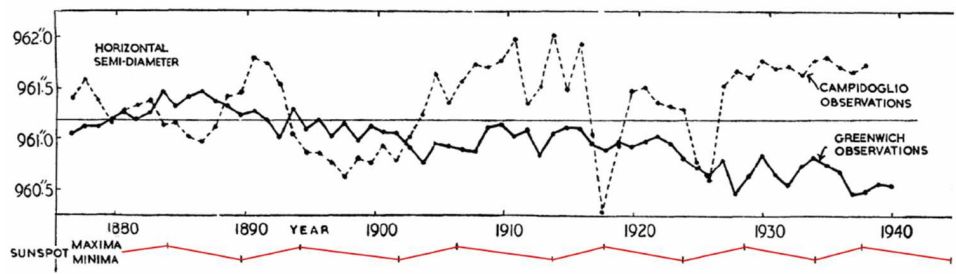}}
\caption{Horizontal semi-diameters of the Sun measured at Greenwich with Airy's meridian circle of 6 inches (15.24 cm). Campidoglio observations, with a 11 cm meridian telescope, are superimposed. The inconsistency between two data sets is evident, from Gething (1955).\protect\cite{Gething}}
\label{Fig. 1}
\end{figure}

\section{Lucernaria Discrete Power Spectrum: First Evidence of Low Frequency Seeing}

The most simple way to measure the seeing is by projection of the solar image on a regular grid during a drift-scan observation. A videocamera records the transit of the solar limbs above the grid, and the time intervals required to cover the evenly spaced intervals of the grid are measured by a frame by frame inspection.
The standard deviation of these time intervals $\sigma$ [s] is related to the seeing $\rho$ [arcsec] by the approximate formula: $\rho = \sigma \cdot 15\cdot \cos(\delta_{\odot})$ where $\delta_{\odot}$ is the declination of the Sun at the moment of the observation.\cite{SigiEgitto2}
The diffraction is the lower limit of the detectable amplitude of the seeing. 
For our application we used the Lucernaria Dome indoors the Basilica Santa Maria degli Angeli e dei Martiri in Rome. The Lucernaria lenses are fixed 6.3 cm solar lenses with focal lenght 20 m.\cite{Cuevas} The resulting diffraction is 2.31 arcsec for $\lambda = 550$ nm. Results in Fig. 2.\cite{NCB}

\begin{figure}
\centerline{\includegraphics[width=1\textwidth,clip=]{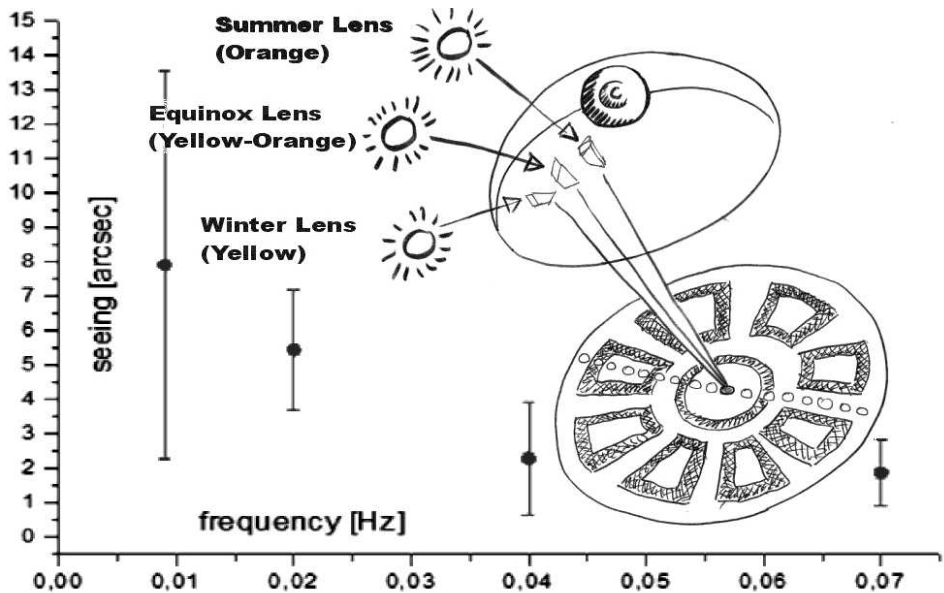}}
\caption{Power spectrum of the seeing at discrete values, measured with the Lucernaria Dome's 6.3 cm solar lenses in the Basilica of Santa Maria degli Angeli, Rome: from the graphic is evident that there is significant power even at low frequencies.}
\label{Fig. 2}
\end{figure}

\section{IRSOL: the Image Motion component of the Seeing}
\subsection{Telescope}
The observation of the full solar disk is performed with Gregory type vacuum telescope 45 cm aperture and 25.0 m focal length.
The instrument gives a portion (about $100 \times 200$ arcsec) of the solar image, from which we recover the curvature of the limb and the solar center.
\subsection{Detector}
The image of the Sun is projected on the CCD Baumer camera, and it is digitalized. The figures here represent the motion of the center of the Sun as recovered from the Northern limb, tracked for 1000 s, and the corresponding FFT power spectrum over frequencies, in abscissa, from 1 to 1/100 of such timespan. The acquisition frequency was 1 image per second. Results in Fig. 3.\cite{NCB,sigi}

\begin{figure}
\centerline{\includegraphics[width=1\textwidth,clip=]{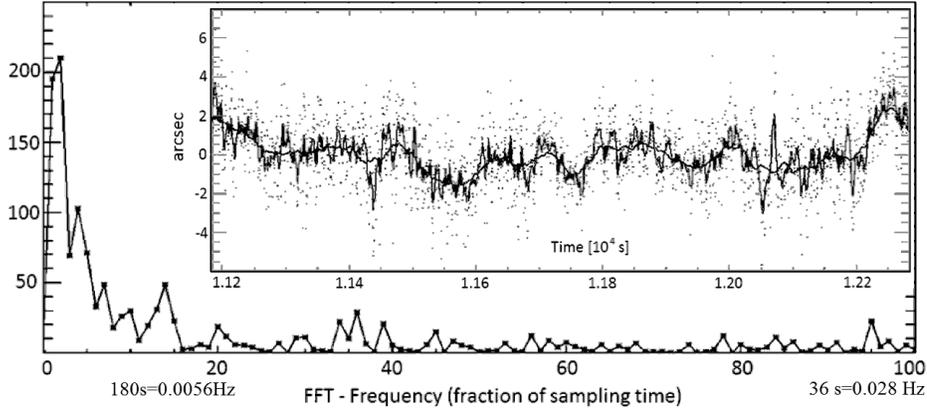}}
\caption{The slow motion of the solar center observed at IRSOL 45 cm telescope. In this image, the effect of ``sub-Hertz" fluctuations is evidenced to the scale of one arcsecond. There is a peak of the power spectrum at 15, i.e. 4 minutes. This can explain the difference between two following measurements of the solar diameter with drift-scan method, and this also explain the need of several measurements to be statistically averaged in order to give a reference value for the solar diameter. But several measurements can occur under different meteorological conditions even in the same day, and the single measurement cannot be considered as statistically independent, as the Gaussian hypothesis requires. That could be one of the reason of the great scatter that the yearly averages published for Greenwich and Capitol Observatory both show.}
\label{Fig. 3}
\end{figure}

\section{Huairou Solar Station: the Streching Component of the Seeing}
\subsection{Telescope}
The observation of the full solar disk is performed with the Solar Magnetism and Activity Telescope (SMAT),\cite{Zhang} that is a telescope with a tele-centric optical system of 10 cm aperture and 77.086 cm effective focal length realized to investigate the global magnetic configuration and the relationship with solar activities. 
The birefringent filter for the measurement of vector magnetic field is centered at 532.419 nm (Fe) and its bandpass is 0.01 nm.
\subsection{Detector}
A CCD camera, Kodak KAI-1020, is used for the measurement of full disk. The image size of the telescope is $7.4 \times 7.4$ mm, and the size of CCD is $992 \times 1004$ pixels. 
The frame rate of the CCD camera is 30 frame $s^{-1}$ and its maximum transmission rate is 60 Mbyte $s^{-1}$. 

\subsection{Full Disk Radial Pulsation Analysis}
The energy found in the power spectrum at longer periods, T=902.4 s, arises from the differential refraction's effect, which changes continuously with zenithal distance. Other peaks of the spectrum are due to the atmospheric seeing. In particular the one at T=300.8 s shows that there is power in the region between 200 s and 400 s, and this confirms the results of IRSOL with a peak at 4 minutes, or 1/15 of the whole period, in Fig. 3.
Another significant period is 19.6 s, i. e. 1/46 of the whole timespan.
 
\begin{figure}
\centerline{\includegraphics[width=1.1\textwidth,clip=]{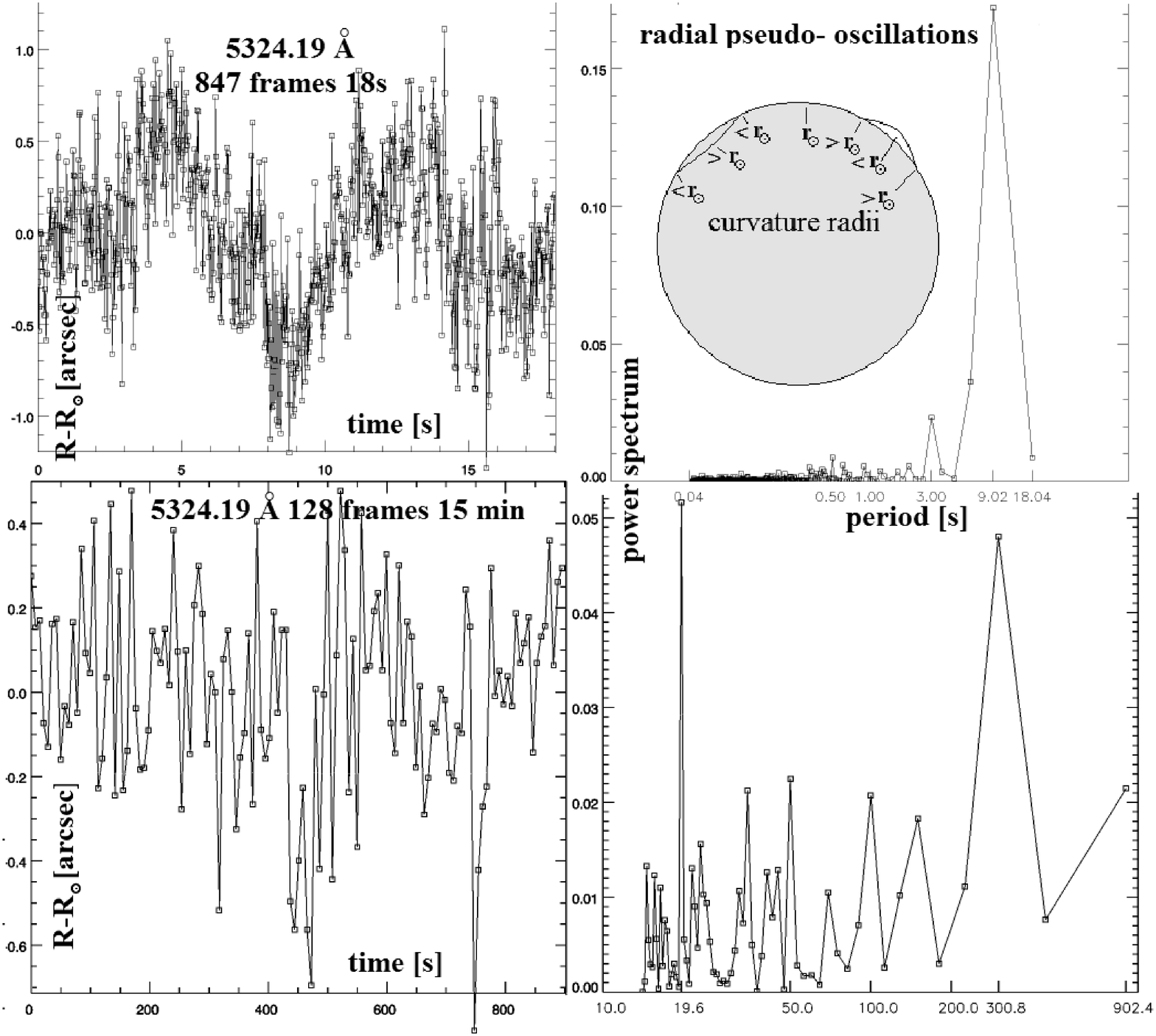}}
\caption{Solar radius instant measurements at Huariou Solar Station SMAT 10 cm telescope. 
Two data collecting mode are performed with 2ms exposure time: 21 ms readout, 847 frames, 18 s (upper level) and 7 s readout, 128 frames, 15 min (lower level).
For every frame of the image we performed a circle fit with the inflexion points' positions.
The fit returns an average value of the radius $R_{obs}$ as well as the position of the center of the Sun $(x_c, y_c)$.
We plot the evolution of $(R_{obs} - R_{\odot})$ as a function of time (left panel), being $R_{\odot}=959.63$ arcsec the standard solar radius at 1 Astronomical Unit: this corresponds to the zero level of the vertical axis. 
We obtain the power spectrum of the fluctuations through a Fast Fourier Transform (right panel). 
Local disturbances inward or outward produce local curvature radii larger or smaller than the solar radius; their changing average produces the pseudo pulsations of the radius.}
\label{Fig. 4}
\end{figure}

\section{Conclusion}
The role of seeing fluctuations between 0.1 Hz and 0.001 Hz is crucial in drift-scan measurements of the solar diameter, either meridian transits or almucantarat transits. This study evidenced this effect in a clear way. 
The fluctuation's scale is not defined here, since we did not apply the analysis to a single point only, but also on solar limb arches of several arcseconds (from 200 arcsec for IRSOL to the whole disk for Huairou Solar Station). Consequently the scale of the seeing (composed by blurring + image deformation + image motion) depends also on the algorithm used to define the solar figure.

An average made with N points distributed over all $360^{\circ}$ of position angles (Huairou Solar Station), is different by the one made over $12^{\circ}$ (IRSOL), and by the one, discrete, made on the preceding and following limb at Santa Maria degli Angeli Lucernaria by visual inspection (about $20^{\circ}$ of the solar limb involved). 
All these measurements have in common the detection of significan energy at the low frequencies regions around 0.01 Hz of the power spectrum. 
These effects have to be taken into consideration for a fruitful analysis of the solar diameter measured with these methods.

A full Sun imager in parallel with the main telescope would help to monitor the whole solar image motion occurred during a single transit observed with the drift-scan method. 
Conversely a similar imager can help to understand the image's stretching phenomena during an heliometric observational session.


\end{document}